# Kinetic Similarity between Extinction Strain Rate and Laminar Flame Speed

Weiqi Ji [a], Tianwei Yang [b,c], Zhuyin Ren [b,c], Sili Deng [a*]

[a] *Department of Mechanical Engineering, Massachusetts of Institute of Technology, MA 02139, USA*

[b] *Center for Combustion Energy, Tsinghua University, Beijing 100084, China*

[c] *School of Aerospace Engineering, Tsinghua University, Beijing 100084, China*

*Preprint*

Correspondence: Sili Deng {silideng@mit.edu}

**Abstract**

Extinction strain rate (ESR) and laminar flame speed (LFS) are fundamental properties of a fuel/air mixture that are often utilized as scaling parameters in turbulent combustion. While LFS at atmospheric and elevated pressures are extensively investigated, experimental measurements of ESR with counterflow premixed flames are very limited for flame instability often occurs near extinction, especially at high pressures. Due to the scarcity of ESR measurements, most combustion kinetic models are mainly validated and optimized against LFS. However, it is questionable whether the controlling reactions are the same for ESR and LFS such that those models are also valid for predicting ESR. This work quantifies the kinetic similarities between ESR and LFS by analyzing their kinetic sensitivity directions. The direction is represented by a unit vector composed of the normalized sensitivity of ESR or LFS to the rate constant for each elemental reaction. Consequently, the similarity between the two directions is measured by the inner product of the corresponding unit vectors. The sensitivity directions of ESR and LFS are found parallel for various fuels, equivalence ratios, and pressures. Furthermore, sensitivity directions at various strain rates are also similar for the maximum temperature, local temperature at various locations in the flame coordinate, and ESR in counterflow premixed flames. These findings suggest that LFS and ESR are similarly effective as the target for constraining and optimizing rate constants in kinetic models. In addition, the independence of the sensitivity directions on the strain rate also enables us to perform uncertainty quantification for turbulent flames with a wide range of strain rates based on the kinetic sensitivity of ESR and LFS.

# 1. Introduction

Flame extinction is a critical combustion performance factor in practical combustion devices. Especially in modern lean-premixed low-emission combustors, where premixed fuel/air mixtures close to lean extinction limits are burned to lower the peak flame temperature and thus NOx formation, flames are susceptible to extinction. The tendency of flame extinction for fuel/air mixtures in strained flows can be quantified by the extinction strain rate (ESR), which has a dimension of the inverse of time. A higher ESR indicates that the mixture is more likely to sustain combustion in strained flows. As a fundamental characteristic of the fuel-oxidizer mixture, similar to the laminar flame speed (LFS), ESR has been used as an important target for the validation of chemical kinetic models. ESR is also one of the fundamental scaling parameters that control the flame structure and stability of turbulent flames in many experimental studies [1–3]. These experiments show that the flame stability correlates well with ESR rather than LFS. Therefore, accurate prediction of ESR by kinetic models is crucial for turbulent combustion modeling.

ESR has been extensively studied in counterflow premixed flames both experimentally and numerically [4–7]. However, ESR has not been studied to the extent of LFS with much fewer experimental datasets than LFS [8]. Particularly, all of the reported ESR of premixed flames are at ambient pressures due to the onset of turbulence at high pressure [9]. Therefore, current combustion kinetic models are mainly validated and optimized against LFS, such as GRI3.0 [10] and FFCM-1 [11], leaving it questionable whether these models are also valid for predicting ESR accurately.

One way to answer this question is to investigate whether the same set of reactions control the flame extinction and propagation. Sensitivity analysis has been widely adopted to study the controlling reactions for LFS and ESR. The sensitivity measures the response of LFS or ESR to the incremental changes of the pre-factor *A* in the kinetic rate constants. Previous studies [6,9,12] have shown that the ESR shares similar sensitive reactions with LFS for various kinds of fuels and equivalence ratios, i.e., the reactions with high sensitivity ranking are almost the same for ESR and LFS. However, besides demonstrating this *qualitative* similarity, little attention has been directed to evaluating the *quantitative* similarity indicated by the sensitivity directions. The sensitivity direction is a unit vector composed of the normalized sensitivity and illustrates the parameter interactions and coupling effects among the sensitive reactions[13]. For example, simultaneously perturbing the rate constants of sensitive reactions along the sensitivity direction will significantly change the model prediction, but retain the same prediction if perturbing along the orthogonal directions. Therefore, a kinetic model that well predicts LFS could fail to predict ESR if the sensitivity directions of LFS and ESR are not parallel.

Therefore, this work investigates the sensitivity directions of ESR and compares them with those of LFS. We shall show that the sensitivity direction of ESR is parallel to that of LFS for various kinds of fuels, equivalence ratios, and pressures. To understand such similarity, we shall further investigate the evolution of the sensitivity directions of flame temperature as the strain rate increases from strong burning states to near-extinction states. In addition, the sensitivity directions of the temperature and species profiles at different locations in the flame coordinate will also be studied. Finally, we shall show that the sensitivity directions for flame temperature are independent of the strain rates and are the same as the sensitivity direction for ESR.

This study will provide insights into the effects of the strain rate on the sensitivity directions and controlling reactions of stretched flames. The identified kinetic similarities between ESR and LFS also facilitate the development of kinetic models and uncertainty quantification for laminar and turbulent flames. Specifically, the significance and implications of the current work are summarized below.

(1) Utilizing the experimental data of ESR to optimize kinetic models can be circumvented. Since the optimization of kinetic models is performed in the sensitivity directions [14], the similarities between the sensitivity directions of ESR and LFS suggest that choosing LFS as the optimization target is sufficient to capture the kinetic response.

(2) Fast sensitivity analysis and uncertainty quantification of ESR become feasible. The kinetic sensitivity of ESR could be efficiently computed based on the sensitivity of LFS, which expedites the computation significantly.

(3) Efficient kinetic uncertainty quantification in turbulent combustion simulations become feasible [15]. The similarity between the sensitivity directions of the ESR and the flame temperature, regardless of the strain rate, suggest that: if the feature of the turbulent flame scales well with ESR [2–4], the sensitivity direction for ESR could also be generalized to represent the sensitivity direction of the entire temperature field.

**2. Kinetic Similarity between Extinction Strain Rate and Laminar Flame Speed**

ESR is usually measured and simulated in twin premixed counterflow flames, the response of which can be described via S-curve analysis. The maximum strain rate that the flame can sustain with fixed composition and thermodynamic states at the inlet is defined as the ESR. Specifically, the absolute value of the maximum velocity gradient is often utilized to evaluate ESR. Numerically, since the ESR corresponds to a singularity point, it has to be solved using either flame continuation approach [16] or reducing factor approach [17].

The continuation approach facilities an efficient sensitivity analysis of the ESR as first proposed in [18]. By realizing that ESR becomes a dependent variable when the two-point continuation approach [16] is invoked, it is possible to perform rigorous sensitivity analysis with respect to kinetic rate constants for ESR at the exact strain rate determined experimentally. In the reducing factor approach, the simulation of the counterflow premixed flame starts from a low strain rate and gradually increases to the near-extinction states. The step size of the increment of the strain rate should be gradually decreased to retain the burning solutions up to the extinction point. Such an approach has been implemented in the open-source software of Ember [17], and the reducing factor approach with Ember has been reported to be more efficient than the continuation approach implemented in Chemkin [19]. However, the sensitivity of ESR has to be evaluated using the finite difference in the reducing factor approach, which could be computationally expensive. Nonetheless, in both approaches, the sensitivity is usually reported in its dimensionless form,

$$s_i = \frac{\partial ln[ESR]}{\partial ln[k_i]} = \frac{k_i}{ESR}\frac{\partial [ESR]}{\partial [k_i]}, \quad (1)$$

where $k_i$ is the rate constant for the $i$-th reaction, and $s_i$ is the corresponding sensitivity coefficient. The sensitivity coefficients for all reactions can be stacked into a sensitivity vector, i.e., $\mathbf{S} = [s_1, s_2, ..., s_n]$. Furthermore, the sensitivity vector is normalized by its magnitude $\|\mathbf{S}\|_2$ to obtain the sensitivity direction $\tilde{\mathbf{S}}$ ($\tilde{\mathbf{S}} = \mathbf{S}/\|\mathbf{S}\|_2$). Therefore, the sensitivity direction represents the normalized sensitivities of all reactions.

Figure 1 shows the sensitivity analysis of LFS and ESR of a methane/air mixture using the detailed mechanism in Hashemi et al. [20]. The equivalence ratio is 0.7, and the pressure is 30 atm, which are kept the same as Long et al. [9]. The unnormalized and normalized sensitivities are shown in Figs. 1a and 1b, respectively. It can be seen that the sensitivity rankings for ESR and LFS are almost the same, with the most sensitive reaction being the primary chain-branching reaction of H+$O_2$<=>O+OH. Overall, the magnitude of the sensitivity of ESR is larger than that of LFS. As pointed out in [9], this result is consistent with previous studies that showed LFS to scale with the square root of the overall reaction rate [21] while ESR was shown to scale linearly [22]. Therefore, normalization of the sensitivity by the sensitivity magnitude is necessary to assess the relative sensitivity of different reactions. Surprisingly, the normalized sensitivity for ESR and LFS are very similar to each other as shown in Fig.1b.

To measure the similarity between two sensitivity directions, the inner product between two normalized sensitivity vectors can be quantified, which is also termed as the cosine similarity [13,23]. The cosine similarity of unity corresponds to a perfect parallel, while zero indicates that two directions are orthogonal to each other. For the normalized sensitivities in Fig. 1b, we evaluated the cosine similarity $\langle \tilde{S}_{ESR}, \tilde{S}_{LFS} \rangle$ to be 0.98, which suggests that the two sensitivity directions are almost parallel. This result implies that the ESR and LFS for the specified mechanism and conditions are kinetically similar, in the sense that they not only share the same set of controlling reactions but also the same response to the coupling among these reactions.

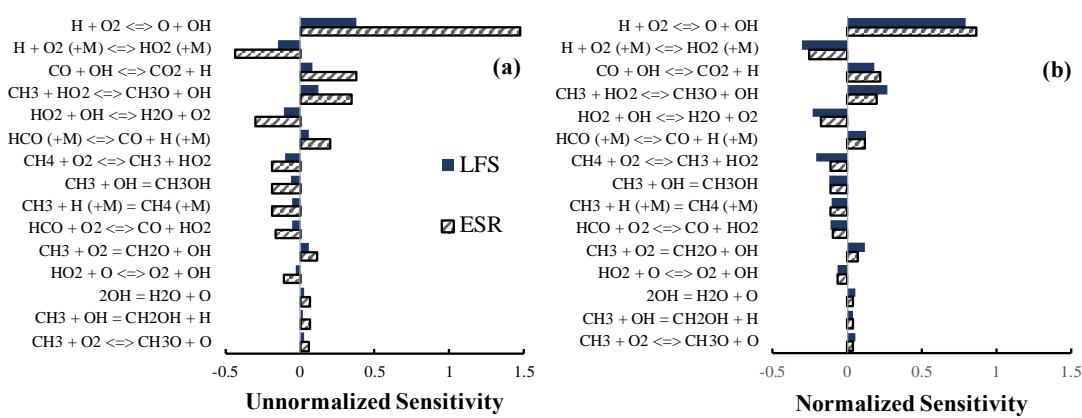

Figure 1. Sensitivity analysis of LFS and ESR of the methane/air mixture with an equivalence ratio of 0.7 and at 30 atm using a detailed mechanism [20]. (a) Unnormalized sensitivities, (b) Normalized sensitivities.

Next, we shall investigate the kinetic similarities between ESR and LFS with other fuels, equivalence ratios, and pressures. Computing the sensitivity of ESR for large hydrocarbon fuels could be time-consuming since computing the ESR itself for large reaction model is already time-consuming [17] and hyperparameter tuning is often needed to save computational time while retaining a stable solution. Fortunately, the kinetic sensitivities of ESR are usually reported in the literature in addition to the experimental measurement of ESR. Therefore, we compiled the sensitivity analysis results of ESR as well as LFS from literature and normalized the sensitivities to obtain the cosine similarity.

The compiled database is summarized in Table 1. The investigated conditions cover a wide variety of fuel, equivalence ratios (phi), and pressures (p [atm]). The "top N" in the table refers to the number of top-ranked sensitive reactions presented in the literature. For most of the cases, the cosine similarity is higher than 0.95, which suggests that the two sensitivity directions for ESR and LFS are very similar to each other. The consistency between large hydrocarbon fuels and methane agrees with previous studies [6,7,12] which suggest that the ESR is most sensitive to small-molecule reactions, i.e., $H_2/CO/C_1$-$C_2$ kinetics.

Note that there are a few outliers that have noticeable lower cosine similarity than the others. Regarding the outliers of #9-10, the difference is potentially because the ESR at lean condition is insensitive to the reaction of $CO+OH<=>CO_2+H$ with the kinetic model published in 2009 [24]. The model significantly underpredicted the ESR for lean and stoichiometric conditions. The predictions with the updated model employed in [25] agreed much better with the experimental data. Correspondingly, the difference in the sensitivity direction is significantly reduced, as shown in #12-13. For the dataset of #2, the cause of the difference is suspected to be that the ESR is extremely sensitive to $C_3H_4+O<=>CH_2O+C_2H_2$ based on the kinetic model published in 1998 [12]. The model showed large discrepancies from the experimental data of ESR. Therefore, it can be summarized that the similarity between the kinetic sensitivity of ESR and LFS is an inherent property of fuel-air mixtures, although such similarity might not show in computations if the models significantly off predict the experimental measurements.

Table 1. The kinetic similarity between ESR and LFS evaluated for the sensitivities in the literature.

| No. | fuel | Phi | cosine | top N | p[atm] | ref. | alias |
|---|---|---|---|---|---|---|---|
| 1 | methanol | 0.77 | 1.00 | 16 | 1 | [5] | Holley2006 |
| 2 | *iso*-octane | 0.87 | 0.46 | 12 | 1 | [5] | Holley2006 |
| 3 | *n*-dodecane | 1.0 | 0.94 | 10 | 1 | [7] | Kumar2007 |
| 4 | *n*-dodecane | 1.0 | 0.97 | 10 | 1 | [7] | Kumar2007 |
| 5 | *n*-dodecane | 1.0 | 0.95 | 10 | 1 | [7] | Kumar2007 |
| 6 | dimethyl ether | 0.6 | 1.00 | 10 | 1 | [28] | Wang2009 |
| 7 | dimethyl ether | 1.0 | 0.99 | 10 | 1 | [28] | Wang2009 |
| 8 | dimethyl ether | 1.5 | 0.91 | 10 | 1 | [28] | Wang2009 |
| 9 | *n*-butanol | 0.6 | 0.59 | 8 | 1 | [25] | Veloo2010 |
| 10 | *n*-butanol | 1.0 | 0.89 | 8 | 1 | [25] | Veloo2010 |
| 11 | *n*-butanol | 1.5 | 0.96 | 8 | 1 | [25] | Veloo2010 |
| 12 | *n*-butanol | 0.6 | 0.95 | 8 | 1 | [25] | Veloo2010 |
| 13 | *n*-butanol | 1.0 | 0.97 | 8 | 1 | [25] | Veloo2010 |
| 14 | *n*-butanol | 1.5 | 0.97 | 8 | 1 | [25] | Veloo2010 |
| 15 | *n*-dodecane | 0.7 | 0.96 | 10 | 1 | [6] | Ji2010 |
| 16 | *n*-dodecane | 1.0 | 0.97 | 10 | 1 | [6] | Ji2010 |
| 17 | *n*-dodecane | 1.4 | 0.97 | 10 | 1 | [6] | Ji2010 |
| 18 | *n*-dodecane | 1.0 | 0.95 | 12 | 1 | [29] | Kumar2010 |
| 19 | *n*-dodecane | 1.0 | 0.97 | 10 | 1 | [29] | Kumar2010 |
| 20 | methyl-butanoate | 0.7 | 0.97 | 14 | 1 | [30] | Wang2011 |
| 21 | methyl-butanoate | 1.0 | 0.98 | 14 | 1 | [30] | Wang2011 |
| 22 | methyl-butanoate | 1.5 | 0.97 | 14 | 1 | [30] | Wang2011 |
| 23 | *n*-propylbenzene | 1.0 | 0.96 | 10 | 1 | [31] | Hui2012 |
| 24 | methane/$C_2HF_5$ | 1.0 | 0.96 | 14 | 1 | [32] | Xu2017 |
| 25 | methane/$C_2HF_3Cl_2$ | 1.0 | 0.99 | 14 | 1 | [32] | Xu2017 |
| 26 | methane/$C_3H_2F_3Br$ | 1.0 | 0.97 | 14 | 1 | [32] | Xu2017 |
| 27 | methane | 0.7 | 0.98 | 15 | 1 | [9] | Long2019 |
| 28 | methane | 0.7 | 0.99 | 15 | 7 | [9] | Long2019 |
| 29 | methane | 0.7 | 0.98 | 15 | 30 | [9] | Long2019 |
| 30 | methane | 1.3 | 0.99 | 15 | 1 | [9] | Long2019 |
| 31 | methane | 1.3 | 0.97 | 15 | 7 | [9] | Long2019 |
| 32 | methane | 1.3 | 0.98 | 15 | 30 | [9] | Long2019 |

It is worth to mention that the kinetic similarity is consistent with the historical views on the connection between ESR and LFS based on the time scale analysis by Peters [26]. The flame residence time in flame propagation is given by the ratio between the flame thickness and LFS, while the quench time for flame extinction is given by the inverse of ESR. The numerical simulations by Rogg [27] for a stoichiometric methane-air flame shows that the flame residence time and quench time are very close to each other, such that the quench time is subsequently approximated by the flame residence time in flame propagation for turbulent combustion modeling. Since both time scales are equal to the chemical time scale [26], one can imply that flame extinction has similar chemical time scale with the unstretched flame propagation, although the flame temperature at extinction can be much lower than the unstretched flame. Since the chemical time scale is determined by the controlling reactions, the similarity of the controlling reactions and sensitivity directions is consistent with the similarity in the time scales.

Leveraging the kinetic similarity between ESR and LFS, the sensitivity of ESR can be efficiently computed based on the sensitivity of LFS. Specifically, the similarity implies that

$$\frac{\partial \ln[ESR]}{\partial \ln[k_i]} \Big/ \frac{\partial \ln[ESR]}{\partial \ln[k_m]} = \frac{\partial \ln[LFS]}{\partial \ln[k_i]} \Big/ \frac{\partial \ln[LFS]}{\partial \ln[k_m]}, \quad (2)$$

where the sensitivity of LFS to all reactions can be efficiently computed with the adjoint approach. Therefore, to evaluate the sensitivity coefficients of ESR for all reactions, we just need to compute the sensitivity of ESR to the most sensitive reaction $k_m$ via the finite difference approach, and sensitivities to other reactions are readily available via Eq. (2). The efficient computation of the sensitivity of ESR will benefit the gradient-based optimization and uncertainty quantification for ESR.

## 3. Kinetic Self-similarity in Counterflow Premixed Flames

The evolution of sensitivity directions for flame temperatures and species concentrations with strain rates are investigated to elucidate the effects of flow strain on the kinetic sensitivity. The sensitivities for temperature and species concentration are also computed using Ember with the finite difference approach for the top 15 reactions pre-selected based on the sensitivity of LFS. The simulation is conducted for a methane/air mixture using the detailed mechanism of GRI3.0 [10]. Details on the composition and thermodynamic states are: equivalence ratio of 0.7, inlet temperatures of 300 K, and atmospheric pressure.

Figure 2a shows the response of the maximum temperature with the strain rate in the upper branch of the S-curve, and the ESR is around 1187 1/s. Since the perturbation of the kinetic rate constants also affects the ESR, the temperature and species sensitivity are computed from 500 to 1075 1/s. Figure 2b then shows the evolution of the sensitivity magnitude and direction for the maximum temperature $T_{\max}$ with strain rates. As expected, the magnitude increases with the strain rates since the flame is more sensitive to chemical kinetics when approaching extinction. Moreover, the evolution of the sensitivity direction is quantified by the cosine similarity evaluated based on the sensitivity direction of ESR. It is shown that the sensitivity directions of $T_{\max}$ are independent of the strain rates, and they are similar to the sensitivity direction of ESR with the minimum cosine similarity of 0.97 for the specified range of strain rates.

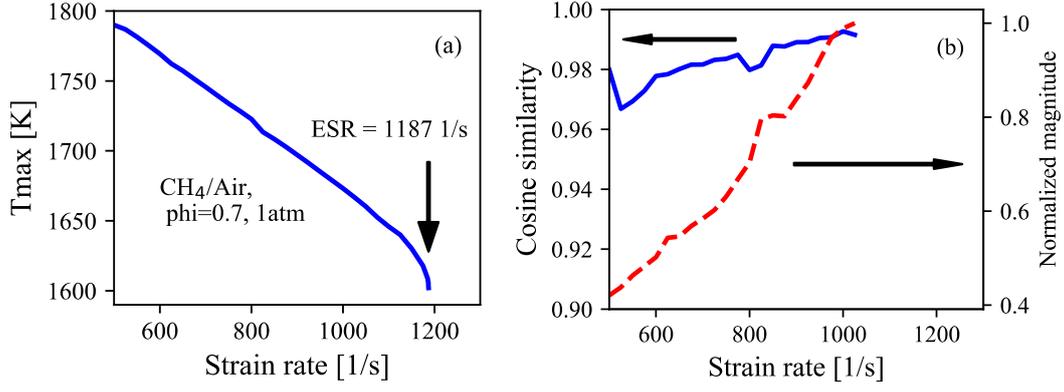

Figure 2. (a) The response of the maximum flame temperature versus the strain rate. (b) The evolution of the cosine similarity and normalized magnitude with strain rates.

The evolution of sensitivity directions for flame temperatures and species concentrations with the locations in the flame coordinate is further investigated to elucidate if there is a universal kinetic sensitivity across the flame. Note that the location in the progress variable space rather than the physical location is shown for better visualization since the compositions change sharply in the narrow reaction zone. The progress variable $C$ is defined in terms of the temperature, i.e.,

$$C = \frac{T - T_{\min}}{T_{\max}(a) - T_{\min}}, \qquad (3)$$

where the maximum temperature $T_{\max}$ changes with the strain rate $a$. The progress variable $C$ is within the range of [0, 1]. The minimum temperature $T_{\min}$ is the inlet temperature of 300 K. Figure 3 shows the evolution of sensitivity directions for temperature and the mass fraction of $CH_4$, CO, H with the progress variable at two representative strain rates. The cosine similarity is evaluated with respect to the sensitivity direction for ESR.

For the temperature sensitivity shown in Fig. 3a, the cosine similarity with ESR is higher than 0.9 in most of the high-temperature regime. Such global similarity has been previously observed in the homogenous auto-ignition system and one-dimensional burner stabilized flame [13]. Furthermore, the temperature sensitivity direction is also approximately independent of the strain rates by comparing the evolution at two distinct strain rates. These observations suggest that the temperature sensitivity direction keeps the same across the flame in the flow with various degrees of strain. Intuitively, the temperature sensitivity is a result of the sensitivity of heat release rate, and the changes of the flame temperature due to flame stretch does not alter the controlling reactions for heat release.

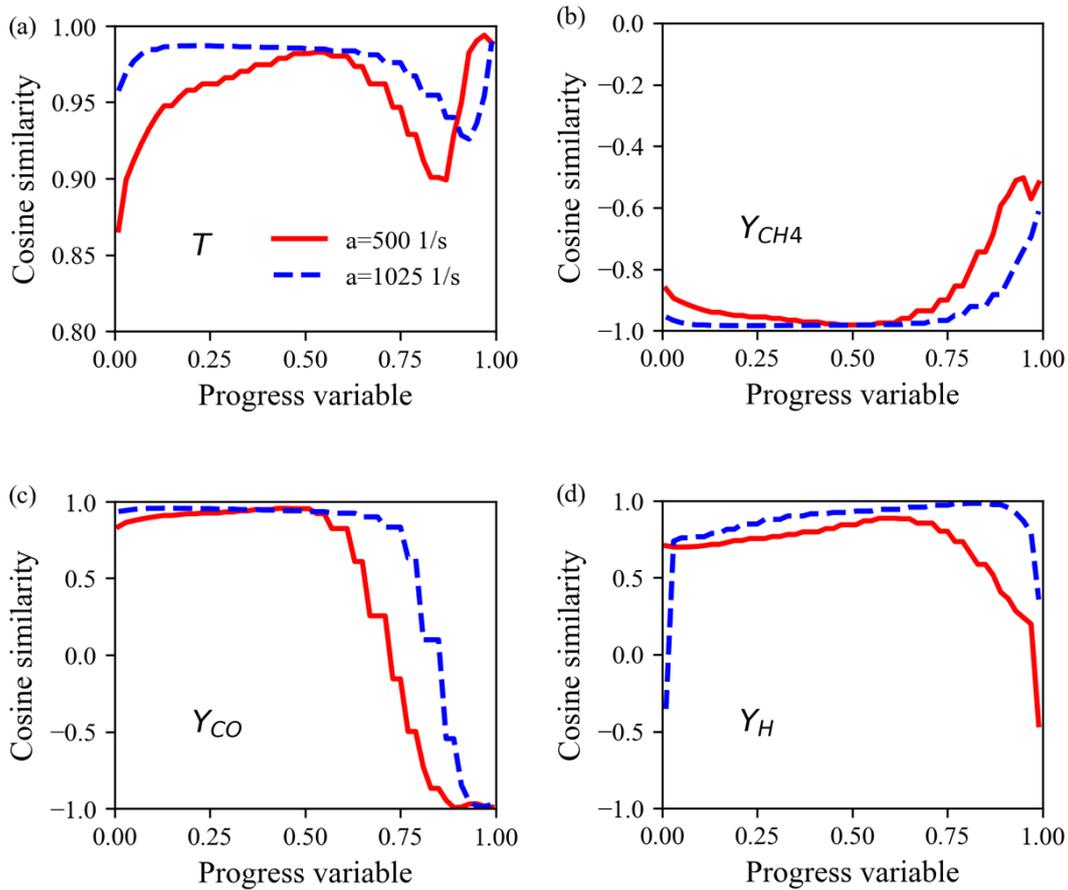

Figure 3. The evolution of the sensitivity direction of temperature and species concentration with progress variable and strain rate (ESR = 1187 1/s). The cosine similarity is computed with respect to the sensitivity direction of ESR.

Similar independence of the strain rate has been previously observed in the sensitivity directions of OH concentration in the simulation of a turbulent lifted flame [33]. Therefore, the independence suggests that the sensitivity direction of ESR might be able to represent the sensitivity direction of the entire temperature field for turbulent premixed flames although a wide range of strain rates or scalar dissipation rates are involved in the flow field. Such representation could substantially accelerate the uncertainty quantification of turbulent combustion simulations using the sensitivity-based method [15].

Finally, the evolution of the sensitivity direction of the species mass fractions with progress variable and strain rate are also studied for their relevance to emissions. The sensitivity directions for major species, $CH_4$ and CO (Figs. 3b and 3c, respectively), are also largely independent of the progress variable and are similar to the sensitivity direction of ESR. The sharp switching between positive and negative correlations corresponds to the local maximum species concentration [13]. The kinetic similarity is much weaker for the mass fraction of H radical compared to the temperature. The weaker similarity could be attributed to the diffusion process [13].

**Conclusions**

We have shown that the kinetic sensitivity direction of ESR is similar to that of LFS for various kinds of fuels, equivalence ratios, and pressures. The similarity is consistent with the flame time scale analysis which shows that flame extinction and propagation share similar chemical time scale although the flame temperature at extinction is much lower. The evolutions of the sensitivity directions for flame temperatures were also studied to explain the kinetic similarity. It is shown that the sensitivity direction of temperature is almost independent of the locations in the flame coordinate and the strain rates, and the temperature sensitivity directions are similar to that of ESR. Therefore, the kinetic similarity could be attributed to the fact that the controlling reactions that determine the temperature sensitivity and ESR are not altered by the maximum flame temperature. The identified kinetic similarity and the independence of strain rates provide insights to the utilization of kinetic models validated and optimized for unstretched and moderate strained flames for practical highly strained flames. These findings also provide foundations for identifying unified sensitivity directions in complex turbulent flames involving a wide range of strain rates and local extinction, to facilitate computational flame diagnostics and uncertainty quantification in turbulent combustion simulations.

**Acknowledgments**

SD would like to acknowledge the support from the d'Arbeloff Career Development allowance at Massachusetts Institute of Technology. TY and ZR were supported by the National Natural Science Foundation of China 91841302.